\input epsf
\documentstyle[pre,aps]{revtex}
\begin{document}

\twocolumn[
\hsize\textwidth\columnwidth\hsize\csname @twocolumnfalse\endcsname

\date{\today}
\title{Reaction Zones and
Quenched Charged-Particle Systems with Long-Range Interactions}
\author{A.D. Rutenberg}
\address{
Centre for the Physics of Materials, Physics Department, McGill University, \\
3600 rue University, Montr\'{e}al QC, Canada H3A 2T8}
\maketitle

\widetext

\begin{abstract}
We determine the evolving segregated or mixed morphology of 
charged-particle systems 
with long-range power-law interactions and overall charge neutrality
that have been quenched to a low temperature. 
Segregated morphology systems are characterized 
by the size of uniformly charged domains, $L(t)$, the particle separation within
the domains, $l_{AA}(t)$, the particle flux-density leaving the domains, $J(t)$,
the width of reaction zones between domains, $W(t)$, the particle
spacing within the reaction-zones, $l_{AB}(t)$, and the particle lifetime
in the reaction-zones, $\tau(t)$.  Mixed morphology systems are essentially
one large reaction zone, with  $L \sim l_{AB} \sim l_{AA}$. By relating
these quantities through the scaling behavior of particle fluxes and
microscopic annihilation rates within reaction-zones, we determine the 
characteristic time-exponents of these quantities at late times.  The
morphology of the system, segregated or mixed, is also determined 
self-consistently.
With this unified approach, we consider systems with diffusion and/or 
long-range interactions, and with either uncorrelated or
correlated high-temperature initial conditions.  
Finally, we discuss 
systems with particle-like topological defects and  electronic
systems in various substrate dimensions --- including quantum-hall devices
with skyrmions. 
\end{abstract}
\narrowtext ] 
\section{Introduction}

Reaction-diffusion systems without long-range interactions 
\cite{Ovchinnikov78,Bramson91,Lee94,Galfi88,Cornell93,Bennaim92,Leyvraz91,%
Howard95}
involve two diffusing species of 
particles that annihilate on contact, $A+B \rightarrow \emptyset$.
The temperature is assumed to be 
low enough that the reverse reaction can be ignored, so
that the densities at late times are determined by the uncorrelated 
initial conditions.  For equal initial mean-densities of the 
particles and anti-particles, there 
are two possible morphologies at late times \cite{Ovchinnikov78}. The first,
for spatial dimension $d \geq 4$, is 
a well mixed morphology in which mean-field dynamics
applies to the particle density, $\partial_t{\bar{\rho}} = -C \bar{\rho}^2$.
This leads to $\bar\rho(t) \sim 1/t$ at late times. The second, for $d<4$,  
is a segregated morphology consisting of single-species domains of
characteristic size $L(t)$. The evolution is 
via diffusive currents feeding particles from the domains into 
reaction zones where they are annihilated by anti-particles. 
The evolution of the mean domain density, 
$\bar\rho \sim t^{-d/4}$, is easily demonstrated when the 
two species have equal diffusion constants, so that the density difference
$\Delta \rho \equiv \rho_A-\rho_B$ satisfies a linear diffusion equation
\cite{Ovchinnikov78,Bramson91}. It also applies more generally
\cite{Lee94}.

In the mixed morphology initial fluctuations decay faster than the 
mean-density and can be ignored. 
In the segregated morphology, long-wavelength initial fluctuations decay
more slowly and asymptotically determine the domain structure. 
In that case, the profile of the domains and the structure
of the reaction-zone can be understood through the particle fluxes and the 
annihilation kinetics of particle pairs 
\cite{Galfi88,Cornell93,Bennaim92,Leyvraz91}, 
as well as on a more formal level \cite{Lee94}. 

It has been recognized from the beginning \cite{Ovchinnikov78} that long-range
interactions between oppositely-charged particles and antiparticles 
change the evolution of a reaction-diffusion system --- both by changing
the initial charge-density 
fluctuations, and by changing the subsequent dynamics. However,
progress has been gradual due to 
the greater complexity of long-range interactions both in analytical
models \cite{propagator}
and in computer simulations, and by the need to understand
the force-free case first.  Nevertheless, progress in long-range models 
has been made along a number of fronts 
\cite{Toyoki88,Ispolatov96,Ginzburg97,Burlatsky96,Ohtsuki84,%
Ginzburg95,Jang95,Zumofen96,Deem98}.
For systems with long-range interactions but no diffusion, and with 
uncorrelated initial conditions, Toyoki presented an analysis of a mixed 
morphology \cite{Toyoki88}. 
Complementing that work, Ispolatov and Krapivsky \cite{Ispolatov96} considered
segregated systems and simulated a variety of force-laws in $d=1$. 
For both long-range interactions and diffusion, with uncorrelated initial
conditions, a self-consistent model by Ginzburg {\em et al.}
\cite{Ginzburg97} and a complementary scaling model \cite{Burlatsky96} has
captured the density evolution for $n \geq d-1$, where $n$ characterizes the
force between particles via $f \sim r^{-n}$ (see Eqn.~(\ref{EQN:force}) below).
The $n=d-1$ Coulombic case has also been treated by Ohtsuki \cite{Ohtsuki84}. 
High-temperature correlated initial conditions
have also been discussed \cite{Ginzburg95}.

In this paper, we are interested in the late-time evolution of initially random 
distributions of charges in homogeneous systems with overall charge neutrality. 
We use the scaling behavior, with respect to length-scale, of the initial 
charge fluctuations, the resulting large-scale currents, and the 
local annihilation rate of particles to determine the morphology as well
as the time-exponents of the characteristic lengths. 
Our approach is based on the assumption that domain
structures are characterized by only two lengths: their size $L(t)$ and their
characteristic particle separation, $l_{AA}(t)$, where the average density
$\bar\rho \sim l_{AA}^{-d}$.  This is comparable to the dynamical-scaling
assumption of phase-ordering systems \cite{Rutenberg95}
and, with appropriate physical input,
leads to a self-consistent description of the system evolution.
Between the domains, we allow for reaction zones of width $W(t)$,
within which both particle species are mixed with typical spacing
$l_{AB}(t)$ and lifetime $\tau$. For segregated systems,
charged currents are absorbed in the reaction zones. For mixed systems, the
reaction-zone pervades the system.  Our approach of balancing currents
with annihilation within reaction zones is in the
spirit of the treatment of force-free systems by 
Redner and Leyvraz \cite{Leyvraz91}; and for these systems 
our results agree with the Renormalization-Group analysis of 
Lee and Cardy \cite{Lee94}.

We ignore correlations apart from the characteristic
lengths $l_{AA}$, $l_{AB}$, $L$, and $W$ which determine the densities
and sizes of the domains and reaction zones. As a result, our
approach is insensitive to early-time dynamics, unequal diffusion
constants \cite{immobile}, and motion of domain boundaries.
All of these factors help determine the {\em amplitude} 
of the growth laws.  Nonetheless, from the scaling properties of 
currents and lengths we can extract the asymptotic 
time-exponents. One benefit of our approach is that
its conclusions are robust to the details of the system, and that it 
provides a vividly physical picture of the morphological evolution of 
charged systems with long-range interactions.

We should emphasize what is new in this paper.
We do not assume which morphology the system selects, but determine
it self-consistently from the physics.  We treat a large number of 
different cases with the same unified approach, 
{\em including} the well understood diffusion-only systems.   We
determine the reaction zone width and density, and also discover
non-trivial domain-edge profiles.
For uncorrelated initial conditions, we discover several regimes where 
ballistic annihilation is the dominant annihilation mechanism.
For high-temperature equilibrium initial conditions, 
we determine the appropriate initial charge-fluctuation spectrum and
show how it modifies the subsequent evolution and morphology of the system.  

\begin{figure}
\begin{center}
\mbox{
\epsfxsize=5.0cm
\epsfbox{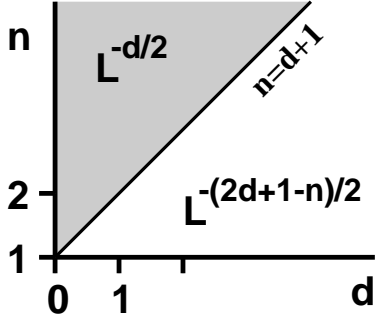}}
\end{center}
\caption{
Charge density fluctuations $\delta \rho \sim L^{-\mu}$ 
for equilibrated high-temperature initial conditions.
In the shaded region the initial conditions
are uncorrelated at large scales; in the unshaded region interactions
reduce large-scale charge fluctuations. 
\label{FIG:initial}}
\end{figure}

\begin{figure}
\begin{center}
\mbox{
\epsfxsize=5.0cm
\epsfbox{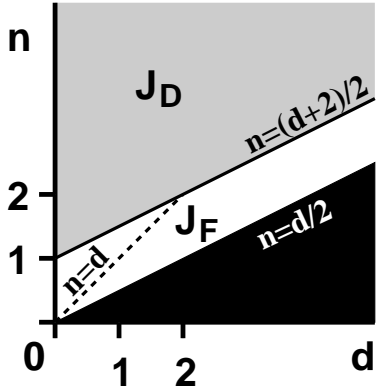}}
\end{center}
\caption{
The asymptotically dominant flux starting from 
uncorrelated initial conditions ($\mu=d/2$). When $J_F$ dominates,
see Tab.~\protect\ref{TAB:F}, when $J_D$ dominates see
Eqn.~(\ref{EQN:JD}). For equilibrated high-temperature initial conditions $J_D$
always dominates at late times. 
\label{FIG:ranflux}}
\end{figure}

\begin{figure}
\begin{center}
\mbox{
\epsfxsize=9.0cm
\epsfbox{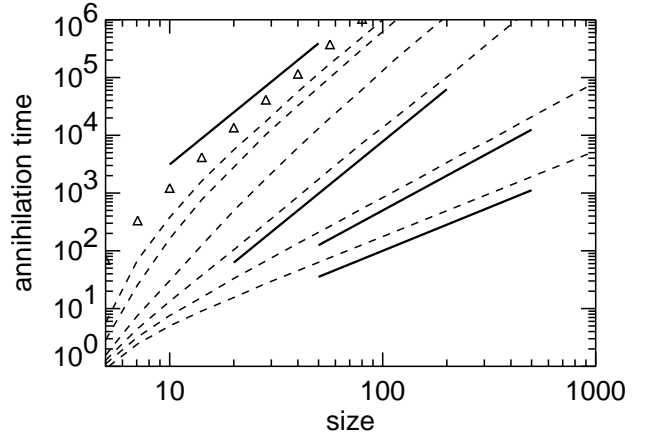}}
\end{center}
\caption{Annihilation time $\tau(l)$ vs size $l$ 
of a particle/anti-particle pair placed in periodic boxes of size $l$ 
in $d=3$, where the particles are initially $l/4$ apart. 
They evolve according to Eqn.~(\protect\ref{EQN:velocity})
with a central force $f = 1/r^n$, mobility $\eta=1$, 
diffusion constant $D=1/4$, and time-step $\delta t=1$. 
The dashed curves correspond to, from bottom to top,
$n=1/2$, $1$, $3/2$, $2$, $3$, and $4$. Triangles correspond
to purely diffusive motion.  The short solid lines indicate the 
expected asymptotic behavior, with $\tau(l) \sim l^{3/2}$,
$l^2$, and $l^3$ (twice), from bottom to top respectively. Diffusion, with
$\tau_D \sim l^3$, dominates asymptotically for $n \geq 1$. 
\label{FIG:ann3}}
\end{figure}

\begin{figure}
\begin{center}
\mbox{
\epsfxsize=5.0cm
\epsfbox{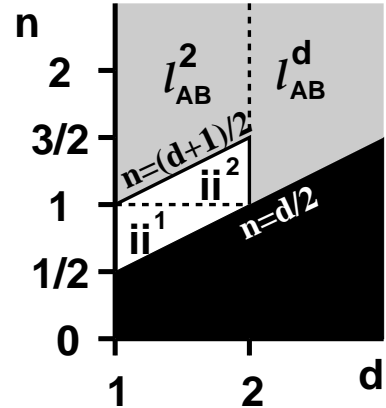}}
\end{center}
\caption{Dominant annihilation times for force and noise, 
with uncorrelated initial conditions. 
Diffusive annihilation ($\tau_D$) dominates in the shaded region, while 
ballistic annihilation ($\tau_B \sim l_{AB}^d F^{d-2}$) 
dominates in the clear region. The regions
$[ii^1]$ and $[ii^2]$ correspond to Fig.~\protect\ref{FIG:growth}.
For equilibrium high-temperature initial conditions, when noise is present, 
$\tau_D$ always dominates.  For force-only annihilation, irrespective
of initial conditions, $\tau_B \sim l_{AB}^d F^{(d-n-1)/n}$ 
dominates for $n<d$ while $\tau_F \sim l_{AB}^{n+1}$ dominates for $n>d$. 
\label{FIG:ann}}
\end{figure}

In the next section we introduce long-range interactions with
overdamped dynamics between initially-random particles and anti-particles. 
For equilibrium initial conditions, 
we calculate the expected initial long-wavelength charge fluctuations,
characterized by $\mu$ 
[Sec.~\ref{SEC:initial}, see also Fig.~\ref{FIG:initial}].
From the charge fluctuations, using a scaling form for the domain
profiles near edges, we determine the scaling of charged currents 
[Sec.~\ref{SEC:profile}, see also Tab.~\ref{TAB:F} and
Fig.~\ref{FIG:ranflux}].
We then consider various mechanisms of particle annihilation 
[Sec.~\ref{SEC:annihilation}, see also Figs.~\ref{FIG:ann3} and \ref{FIG:ann}].
We combine our results on currents and on annihilation, by
balancing the currents with the annihilation rates in the 
reaction zones [Sec.~\ref{SEC:morphology}].  
This is sufficient to determine both the system morphology, and the time 
dependence of the domain size $L$ and of the average density, $\bar\rho$
[Fig.~\ref{FIG:growth} and Tab.~\ref{TAB:growth}].
We next consider the reaction zone in more detail, and 
determine its width $W$ and particle spacing $l_{AB}$, as well as
the typical lifetime $\tau$ within the reaction zone
[Sec.~\ref{SEC:W}, see also Tab.~\ref{TAB:tau}].  
Finally we discuss our results, including 
implications for coarse-grained treatments [Sec.~\ref{SEC:discussion}].

We discuss the previous reaction-diffusion literature [Sec.~\ref{SEC:prev}],
then discuss applying our results to
electrically charged ($n=2$) systems in various substrate dimensions
[Sec.~\ref{SEC:electrical}].  We emphasize 
that for $d<3$ novel decay laws and segregated morphologies are found.
We then discuss the application of our results to 
quenched phase-ordering systems with point-like topological defects
[Sec.~\ref{SEC:topo}]. We can also extend
our approach to  include L\'{e}vy super-diffusive systems, as well
as to sub-diffusive systems [Sec.~\ref{SEC:levy}]. 

\begin{figure}
\begin{center}
\mbox{
\epsfxsize=9.0cm
\epsfbox{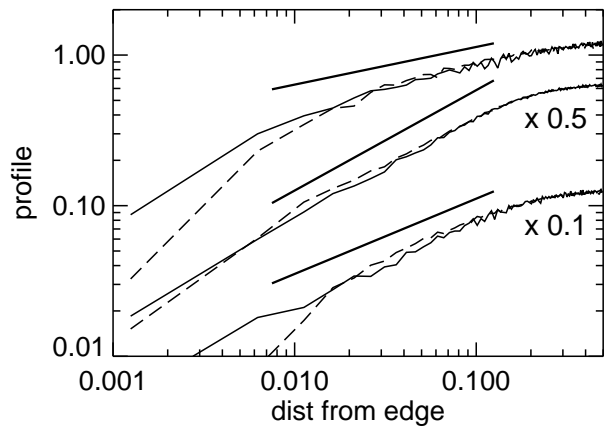}}
\end{center}
\caption{
Average domain profile $f(x)$ vs scaled distance $x$ from
the edge of the domain. 
From top to bottom are $n=3/2$ (at $t=1.9 \times 10^{-5}$ dashed
and $0.0013$ solid), $n=3/2$ ($t=0.0038$ and $0.09$ for dashed and
solid respectively) with
nearest-neighbor interactions, and $n=2$ ($t=1.9 \times 10^{-5}$ and $0.25$
respectively) with nearest-neighbor interactions;
all are in $d=1$.  The second and third sets of curves have been scaled by 
$0.5$ and $0.1$ respectively. The straight segments indicate the expected
domain profiles: $\alpha = 1/4$, $2/3$, and $1/2$.   The crossover at 
small $x$ is due to the reaction zone, and moves to smaller scaled distance
at later times. 
\label{FIG:prof}}
\end{figure}

The effect of short-range cutoffs on the power-law interactions
can easily be treated [Sec.~\ref{SEC:cutoff}].  
We numerically explore our results on domain profiles 
[Sec.~\ref{SEC:numerical}, see also Fig.~\ref{FIG:prof}]. 
Finally, we conclude [Sec.~\ref{SEC:summary}]. 

Throughout this paper we concentrate on exponents, or scaling
dependencies, of various quantities in the late-time limit. Inequalities
apply to exponents, so that a process is dominant if
it is asymptotically largest as $t \rightarrow \infty$. 
We denote the mechanisms with a subscript $D$, $F$, or $B$ for
diffusive, local force driven, or long-range ballistic processes,
respectively.  

\section{Long-Range Interactions}
\label{SEC:longrange}

We work in the overdamped limit, in which particle velocity equals
the applied force times a constant mobility, $\eta$.  The pairwise forces are
\begin{equation}
\label{EQN:force}
		f_{ij} = (n-1) C q_i q_j/ r_{ij}^n,
\end{equation}
between particles with charges $\{ q_i \}$ and pairwise separations 
$\{ r_{ij} \}$. 
This corresponds to a pairwise interaction energy 
$E_{ij} = C q_i q_j r_{ij}^{1-n}$, with a logarithm at $n=1$. 
Thus the particle velocity is 
\begin{equation}
\label{EQN:velocity}
	\partial_t \vec{r}_i 
		= \eta \sum_{j \neq i} f_{ij} \hat{r}_{ij} +\vec{\phi}_i,
\end{equation}
where we have added a random uncorrelated 
noise for diffusive motion with diffusion constant $D$, 
where $\langle \vec{\phi}_i(t) \cdot \vec{\phi}_j(t')
\rangle = D \delta_{ij} \delta(t-t')$, 
When oppositely charged particles get within a fixed capture radius $r_c$, 
they annihilate instantaneously \cite{equil}.
We are interested in the behavior of the system at late times when 
distinct length-scales are well-separated. 

\subsection{Initial Conditions}
\label{SEC:initial}

Most studies of reaction diffusion with long-range
interactions have focused on the case of random uncorrelated 
initial conditions, where the particles are randomly placed with
a local Poisson distribution 
\cite{Toyoki88,Ispolatov96,Ginzburg97,Burlatsky96,Ohtsuki84,Jang95,Deem98,%
Zumofen96,Deem98}.  Experimentally, it is more natural 
to quench a system of charged particles from a
high-temperature state in which they are in thermal equilibrium
\cite{Ovchinnikov78,Ginzburg95}. 

Charge fluctuations may be usefully characterized by 
the typical excess charge density at scale $L$, 
\begin{equation}
\label{EQN:mui}
	\delta \rho \sim L^{-\mu}.
\end{equation}
This can be thought of as the excess charge density in a region of 
size $L$ {\em after} coarse-graining to that scale.  It is also related to the
magnitude of the Fourier-component $\rho_{1/L}$ of the charge-density by 
\begin{equation}
\label{EQN:rhok}
	|\rho_{1/L}| \sim L^{d/2-\mu},
\end{equation}
as obtained by summing up 
$O(L_\infty^d/L^d)$ uncorrelated contributions of size $L^d \delta \rho(L)$ and
normalizing out the constant contribution due to the 
system size, $L_\infty$. As a result 
we can obtain $\mu$ directly from $\rho_k$.

We are only interested in the scale dependence of 
initial charge fluctuations that survive for long times
--- i.e. for fluctuations at large scales and correspondingly small $k$. 
Hence we consider a continuum charge-density $\rho(r)$ with energy 
\begin{eqnarray}
	E &= & C/2 \int d^d r d^d r' \rho(r) \rho(r')/|r-r'|^{n-1} \nonumber \\
	 & = & C/2 \int \frac{d^dk}{(2 \pi)^d} \rho_k \rho_{-k} \epsilon_k.
\label{EQN:E}
\end{eqnarray}
From $\epsilon_k \equiv \int d^dx e^{i k \cdot x}x^{1-n}$, we
obtain $\epsilon_k  = \pi^{d/2} (k/2)^{n-1-d} 
\Gamma\left[(d+1-n)/2\right] /\Gamma\left[(n-1)/2\right]$ for $1<n<d+1$.
For $n \geq d+1$ we must introduce a UV cutoff, the inverse
particle separation \cite{initialdensity}, which determines 
a leading $\epsilon_k = const$ small-$k$ behavior.
For $n=1$ the interaction in Eqn.~(\ref{EQN:E}) should be logarithmic, and 
we get $\epsilon_k = \pi^{d/2} 2^{d-1} k^{-d} \Gamma(d/2)$.
Imposing equipartition on Eqn.~(\ref{EQN:E}) with  temperature $T$, 
we get $\langle \rho_k \rho_{-k} \rangle \sim k_B T / \epsilon_k$. 

This derivation reproduces the small $k$ behavior of a more complete
variational approach, see, e.g. \cite{Chaikin95}. Entropy factors do
not affect the leading long-wavelength fluctuations, so that the
integrand in Eqn.~(\ref{EQN:E}) is correct for small $k$. This means
that $\rho_k$ is Gaussian distributed for $k \rightarrow 0$, and 
$\langle |\rho_k| \rangle \sim \epsilon_k^{-1/2}$. 
Comparing with Eqn.~(\ref{EQN:rhok}) we obtain
\begin{equation} 
\label{EQN:mu}
	\mu = \left\{
                \begin{array}{c}
        		(2d+1-n)/2 \ \ \ \ \ \         1 \leq n< d+1 \\
        	d/2 \ \ \ \ \ \ \ \ \ \ \ \ \ \ \ \ \ \ \ \ \ \ \ \ \ \ \ n> d+1.
                \end{array} \right.	
\end{equation}
This is illustrated in Fig.~\ref{FIG:initial}.
Note that $d/2 \leq \mu \leq d$, so that long-range interactions always
suppress initial charge-fluctuations. 
The maximal suppression is achieved at $n=1$ when $\mu=d$.  [For $n<1$
we expect higher-point correlations to be significant.]
For sharp enough interactions, with $n>d+1$, we recover
uncorrelated initial conditions with $\mu=d/2$.

The charge excess at scale $L$, as discussed above, is quite different 
from the charge excess within a sharp boundary of scale
$L$, for example inside a sphere of radius $L$. 
The later has been proposed as a measure
of charge fluctuations, particularly in Coulombic systems in general
dimensions, with $n=d-1$, where a Gauss's law applies (see, e.g., 
\cite{Ovchinnikov78,Ginzburg95,Jang95}) and in systems with 
topological defects \cite{Einhorn80,Mondello90,Dhar81} where similar
integral identities apply to the topological charge density.
At high temperatures the charge inside a given closed
surface is proportional to the
square-root of the surface area, as obtained from integrating the
appropriate random high-temperature field over the surface
\cite{Dhar81}.  Taken literally
this would imply that $\mu = (d+1)/2$ \cite{Ginzburg95}, i.e. larger 
charge fluctuations than indicated in Eqn.~(\ref{EQN:mu}) for Coulombic
interactions. 
However a sharp surface picks up charge fluctuations at short scales
{\em in addition to} the desired large-scale fluctuations. Indeed, the 
Coulombic or high-temperature nature of the system is moot 
---  arbitrary {\em sharp} surfaces within a system with microscopic 
charge heterogeneities (e.g. atoms or thermal fluctuations) 
pick up charge fluctuations proportional to the square-root of the 
surface area.  The useful and appropriate 
charge fluctuations at scale $L$ are given by Eqn.~(\ref{EQN:mu}),
which are not mixed with charge-fluctuations at short-scales \cite{2dxy}.

\subsection{Domain Profile and Currents}
\label{SEC:profile}

Charge fluctuations, coarse-grained to scales much larger than typical
particle separations, decay via charged currents. These currents
can be diffusive, or can be driven by the long-range interactions. 
We first consider 
segregated systems, where the domain size, $L(t)$, sets the scale of 
surviving charge fluctuations. The average
density within a domain, $\bar\rho \sim l_{AA}^{-d}$ is simply proportional
to the initial fluctuations at the domain scale, 
$\delta \rho(L)$, so that \cite{Ginzburg95}
\begin{equation}
\label{EQN:L}
	L \sim l_{AA}^{d/\mu}.
\end{equation}
Charge fluctuations at scales much larger than $L$ cannot have
relaxed since charge transport at larger scales 
is cut-off by the domain structure. 
We now {\em assume} \cite{fractalprofile} that
the domain profile is only determined by the average density, $\bar{\rho}$, 
and size, $L$, of domains. For example, domains 
have a typical density profile 
\begin{eqnarray}
\label{EQN:domain}
	\rho(r)& = & l_{AA}^{-d} f(r/L), \nonumber \\
			& \sim & l_{AA}^{-d} (r/L)^\alpha,
\end{eqnarray}
where $r$ is measured from the edge of the domain. The
second equation holds near a domain edge, with $r \ll L$.
We also require $r \gg W$ so that the profile is probed
well away from the reaction zone. 
This scaling form has been proposed by 
Leyvraz and Redner in diffusive systems \cite{Leyvraz91}, and  was
numerically confirmed in $d=1$. 

The assumption of an invariant scaled domain morphology is powerful, and is
sufficient to determine the coarse-grained current density $J$ near the domain
edge. The flux must have a dominant non-zero constant contribution 
near the domain edge for $r \ll L$ arising from the evolution of domain
density, since no annihilation takes place in the 
domain interior.  Consider the net charge of a domain, 
$Q \sim L^d \bar\rho \sim L^{d-\mu}$. It implies a non-zero net flux density 
$J \sim \dot{Q}/L^{d-1} \sim L^{1-\mu}/t$ near the domain edge.
The exponent characterizing the domain profile, $\alpha$, 
is constrained to allow the dominant $J$ to be finite but non-zero for
$r/L \rightarrow 0^+$.  

If the dominant current is diffusive, then $J_D \sim \bar{\rho}/L$, so that  
\begin{equation}
\label{EQN:JD}
	J_D \sim L^{-(1+\mu)}.
\end{equation}
Imposing a constant diffusive current condition at the domain edge implies 
a linear profile, with $\alpha=1$. This agrees with studies of domain profiles
in diffusive systems \cite{Leyvraz91}.

If the dominant current is driven by long-range forces then
it is given by the coarse-grained field times the local
charge density, $J_F \sim \rho(r) F(r)$. The field, $F(r)$, at a distance
$r$ away from the domain edge due
to the charge distribution given by Eqn.~(\ref{EQN:domain}) is
\begin{equation}
\label{EQN:F}
		F(r) \sim \int_{l_{AA}(r)}^L d^d x [\rho(r+x)-\rho(r-x)]/x^n.
\end{equation}	
We have used the domain scale as the long-distance cutoff, and 
the local interparticle spacing, $l_{AA}(r) \sim \rho(r)^{-1/d}$, 
as the short-distance cutoff. We also restrict the angular integral
to be close to the normal direction from the interface, which retains the 
scaling behavior of $F(r)$ without requiring detailed information 
about the domain shape. 

For $d>n$, charges at distances of order $L$ determines $F$,
so that $F \sim \bar\rho L^{d-n} \sim L^{d-n-\mu}$. 
For a constant flux $J_F$ at small $r$, we must have $\rho(r) \sim const.$
(i.e. $\alpha=0$)
so that $J_F \sim \bar\rho^2 L^{d-n} \sim L^{d-n-2\mu}$. 
When $d>2n$, with uncorrelated initial conditions ($\mu=d/2$),
$F$ increases with the upper-cutoff of the integral in Eqn.~(\ref{EQN:F}),
so that we should use the {\em system-size}, $L_\infty$, rather than $L$.
For this case, the system size enters the dynamics, 
and the thermodynamic limit does not exist \cite{Toyoki88,Ispolatov96}.

For $d<n$, the local charge distribution dominates the $F(r)$ integral.
We can expand the density for $x \ll r$ and find
$F(r) \sim \rho(r)^{(n-1)/d} r^{-1}$ for $n>d+1$. For $d<n<d+1$, on the 
other hand, the integral is dominated by scales around $r$, and we
find $F(r) \sim r^{\alpha+d-n}/[l_{AA}^d L^\alpha]$. 
Insisting on $J_F$ approaches a non-zero constant near the domain
edge, we obtain Table~\ref{TAB:F}.  These results apply far from the 
reaction zone, but otherwise close to the domain edge: $W \ll r \ll L$. [We
determine the reaction zone width in Sec.~\ref{SEC:W}.] 
We see that currents dominantly driven by power-law interactions lead to
non-trivial domain profiles, with $0 \leq \alpha \leq 1/2$, in dramatic
contrast to the diffusive case where $\alpha =1$. It is also interesting
that the current, $J_F$, has the long-range form for $d<n<d+1$, even though
$\alpha \neq 0$.  


When both long-range forces and diffusive effects are present, then we
must compare $J_F$ and $J_D$, and identify which is larger at late times. 
We summarize the results in Fig.~\ref{FIG:ranflux} for random uncorrelated
initial conditions.  For equilibrated high-temperature initial
conditions $J_D$ is always asymptotically larger at late times.

\subsection{Particle Annihilation}
\label{SEC:annihilation}

In a well mixed region of the system, where the typical spacing
between oppositely charged particles is $l_{AB}$, what is the scaling
of the particle
lifetime $\tau(l_{AB})$? There are three annihilation mechanisms:
diffusive annihilation ($\tau_D$), local interaction-driven annihilation
($\tau_F$), and ballistic annihilation ($\tau_B$). We determine their 
scaling dependence on $l_{AB}$ and hence identify the dominant mechanism
at late times.  Our essentially microscopic approach also provides insight into
the applicability of coarse-grained treatments (see Sec.~\ref{SEC:discussion}).

In the force-free case, particles move diffusively with
diffusion constant $D$, and annihilate with oppositely charged particles when 
they approach within a fixed distance $r_c$.  In $d \leq 2$, trajectories 
are space filling and $\tau_D \sim l_{AB}^2/D$ --- the time it takes for a 
particle to diffuse $l_{AB}$. For $d > 2$, 
$\tau_D \sim l_{AB}^d/ (D r_c^{d-2})$, since each 
particle must explore the characteristic volume per particle to find
an anti-particle to annihilate. We have 
\begin{equation}
\label{EQN:tauD}
	\tau_D \sim  \left\{
                \begin{array}{c}
        		 l_{AB}^2         \ \ \ \ \ \  	d \leq 2, \\
        		 l_{AB}^d            \ \ \ \ \ \ \    d> 2.
                \end{array} \right.	
\end{equation}

In the diffusion-free case, considering only local interactions,
$f \sim r^{-n}$, between two particles 
initially separated by $l_{AB}$, the annihilation time 
\begin{equation}
\label{EQN:tauF}
	\tau_F \sim l_{AB}^{n+1}. 
\end{equation}
For many particles in a region, the same result follows from the
scaling of the velocities in Eqn.~(\ref{EQN:velocity}).

With both diffusion and local interactions, 
diffusion ($\tau_D$) dominates the annihilation time for $n>1$, while 
for $n<1$ the force ($\tau_F$) does.  This follows directly
from the particle dynamics, Eqn.~(\ref{EQN:velocity}). 
Rescale all distances by $l_{AB}$, and rescale time by $l_{AB}^2$, 
so that diffusion is unchanged in the scaled coordinates as $l_{AB}$ increases. 
Scaled velocities due to the force are then multiplied by
$l_{AB}^{1-n}$.  As a result, forces do not asymptotically
contribute for $n>1$, while forces dominate for $n<1$.
For $d >2$, this leads to the initially counter-intuitive
result that diffusion dominates for $1<n<d-1$, even though
$\tau_D \gg \tau_F$. This is a well-known result for $n=2$
\cite{Onsager38}.  Essentially, the competition between diffusion and the 
attractive interaction is along the
separation vector between two particles, and hence is always
one-dimensional in character.  
Indeed, in $d=1$ the faster annihilation mechanism
dominates, and the marginal value is $n=1$. 
We have confirmed these predictions for various $n$
in $d \leq 3$ by placing a particle and an anti-particle 
in a periodic box, and plotting the average annihilation time as a 
function of the box size. In Fig.~\ref{FIG:ann3},
we show our results for $d=3$. For $n>1$, the long-range force 
merely changes the effective capture size, leaving $\tau_D \sim
l_{AB}^d$. 


When an applied or non-local force $F$ is present, we must also consider
ballistic annihilation. 
With a velocity proportional to $F$, and in a time $\tau_B$, 
particles sweep out a volume proportional to $\tau_B F l_\ast^{d-1}$,
where $l_\ast$ is the radius of the effective capture cross-section.
Equating this to the typical volume $l_{AB}^d$ per particle, we obtain 
the typical ballistic annihilation time
\begin{equation}
	\tau_B \sim l_{AB}^d l_\ast^{1-d}/F.
\label{EQN:tauB}
\end{equation}
When no noise is present, a particle will be captured by local interactions
only at separations less than $l_\ast$ where local interactions
are as large as $F$, so that  $F \sim f \sim 1/l_\ast^n$ or  
$l_\ast \sim F^{-1/n}$.  We can use the 
typical forces $F(r)$ from Table~\ref{TAB:F}, but must evaluate them 
at the reaction-zone width $W$ (see below, Sec.~\ref{SEC:W}) --- since
the reaction-zone itself is neutral at a coarse-grained scale. Because
$F$ depends on the system morphology, this requires a self-consistent 
solution \cite{selfconsist}. The results are simple:
for $n>d$ we find 
$l_\ast \gtrsim l_{AB}$ and the dominant time-scale is $\tau_F$, 
while for $n<d$ we find $l_\ast \lesssim l_{AB}$ and 
so the dominant time-scale is $\tau_B$. This corresponds to the naive
phase-space result from Eqn.~(\ref{EQN:F}) that charges from 
far away dominate local forces, and hence $\tau_B$ dominates, only
when $n<d$.

When diffusive noise is present, we first ignore the local interactions and only
consider an applied field $F$.  For $d>2$, both the random walk and
the ballistic force sweep out volume proportional to time. Because $F$
decreases with time, $\tau_D$ always asymptotically dominates. 
However, for $d \leq 2$ random walks recur, and the ballistic
drift can enhance the volume covered by the random walk by suppressing 
the recurrence.  The rate of volume swept out by the drifting particle,
is a sphere of radius $l_{\ast}$ every $\Delta t \sim l_{\ast}/F \sim 
l_\ast^2/D$. This implies $l_\ast \sim 1/F$, and leads to 
$\tau_B \sim l_{AB}^d F^{d-2}$.  

When both noise and local interactions are present, the local capture
cross-section $l_\ast$ is determined by the dominant non-ballistic
process --- i.e. $\tau_D$ for $n>1$ and $\tau_F$ for $n<1$.  
The shortest annihilation time for uncorrelated initial conditions
is given in Fig.~\ref{FIG:ann}.  For equilibrated initial conditions, $\tau_D$ 
always dominates for $n>1$.


Our treatment of particle annihilation is essentially microscopic rather 
than coarse-grained, since we have built the
particle separation, $l_{AB}$, directly into the annihilation times.
We have derived our results by considering particle pairs, though
we have included some multi-particle effects 
by never allowing particles to ``escape'' further than $l_{AB}$ 
from an anti-particle. We apply the results in mixed regions of the 
system such as reaction zones. 

\subsection{Domain Morphology: Segregated or Mixed}
\label{SEC:morphology}

We assume the system is described by one of two morphologies
\cite{othermorphologies}, depending on whether coarse-grained 
charge fluctuations are comparable to or much less than the 
mean particle density at late times.  
The former case describes  a segregated morphology
with domains of particles separate from domains of antiparticles. In the 
second case, there is a mixed morphology with no clearly defined domains.
We can use the dominant flux $J$ and fastest annihilation time $\tau$ 
to see which morphology is consistent. The system turns out to have 
a unique consistent morphology: either mixed or segregated. 

First consider a segregated morphology.  The system has domains
of scale $L$ separated by reaction zones of width $W$. Within the zones
there is a typical particle spacing $l_{AB}$. We ignore correlations
or structure within the zones \cite{nonuniform}. Our self-consistency
constraints are that $l_{AB} \lesssim l_{AA}$, and that $W \lesssim L$.
We impose the former because annihilation takes place in the reaction zone
but not in the domain bulk, so the density in the reaction zone should
be smaller as a result.  We impose the latter since if it were not the case, 
with $W \gg L$, the system would be effectively {\em all} reaction zone and
of a mixed morphology.   These constrain the maximum rate that reaction
zones can ``process'' incoming particles:
the average density of particles that are in reaction zones, 
$W L^{d-1}/ (l_{AB}^d L^d)$, is at most $1/l_{AA}^d$.
Conversely, the maximum annihilation rate is $1/\tau(l_{AA})$,
since $\tau(l)$ is monotonic. 
Combining these, the maximum annihilation rate from reaction zones is 
$\partial_t{\bar\rho}_{max} \sim -1/[l_{AA}^d  \tau(l_{AA})]$. 
For segregated systems, the actual 
rate of particles annihilating per unit volume
is determined by the currents entering the reaction zones: 
$\partial_t{\bar\rho}_{seg} \sim -J L^{d-1}/L^d \sim -J/L$.

If the maximum reaction rate $\partial_t{\bar\rho}_{max}$ 
asymptotically dominates the actual rate 
$\partial_t{\bar\rho}_{seg}$, then a domain structure is consistent. 
Indeed, since $\partial_t{\bar\rho}_{max}$ is the rate of density
decrease in a mixed morphology, the inequality indicates that 
the background density decay quickly with respect to the charge
fluctuations --- i.e. that the segregated structure occurs when
it is consistent.  If the maximum reaction rate
is less than the particle flux, then segregated domains {\em cannot}
be sustained, and the mixed morphology should result. 

For the segregated morphology, we use 
$\partial_t {\bar\rho} \sim -J/L \sim -\bar\rho/t$ 
and $L \sim l_{AA}^{d/\mu}$ to extract the domain scale and 
particle density within the domain.  

For the mixed morphology, the flux does not drive annihilation of the mean
density. Rather every particle is effectively in a reaction zone and has a
characteristic lifetime $\tau$. Comparing this to the scaling of the 
density evolution, $\partial_t \bar\rho \sim - \bar\rho/t \sim -
\bar\rho/\tau$, indicates that $\tau(l_{AA}) \sim t$. 

The resulting growth law regimes are shown in 
Fig.~\ref{FIG:growth} for systems with either uncorrelated or high-temperature
equilibrated initial conditions, and with either force-only or force-and-noise
dynamics. The exponents are summarized in Tab.~\ref{TAB:growth}.  
The noise-only case reproduces the 
Ovchinnikov-Zeldovich-Toussaint-Wilczek result \cite{Ovchinnikov78}, and
is given in Fig.~\ref{FIG:growth} b) 
by the short-range ($n \rightarrow \infty$) limit of the 
force-and-noise dynamics with uncorrelated initial conditions. 



We can now consider the consistency of the mixed morphology.  
The system has no domain structure, and has local charge separation
scales $l_{AB} \sim l_{AA}$. Coarse-grained charge fluctuations, above
a scale $X(t)$, remain from
the initial conditions. The mixed morphology is 
consistent if the charge density of the remaining charge fluctuations
$\delta\rho(X) \sim X^{-\mu}$ 
is much less than the mean particle density $\bar\rho \sim l_{AA}^{-d}$. 

Given the scaling of net currents $J(X)$, the charge fluctuations
evolve with $\partial_t \delta\rho(X) \sim \delta\rho/t \sim J
X^{d-1}/X^d \sim J/X$.  For diffusive currents, 
$J_D(X) \sim \delta\rho(X)/X \sim X^{-(\mu+1)}$, and we find that
$X \sim t^{1/2}$.  For force driven currents, $F(X) \sim X^{d-n-\mu}$
from Eqn.~(\ref{EQN:F}) using $X$ as the upper and lower cutoff.  The
force drives the net charge density $\delta\rho(X)$, and yields
a net current 
$J_F(X) \sim X^{d-n-2 \mu}$, leading to $X \sim t^{1/(\mu+n+1-d)}$.
Using the results for $l_{AA}$ from above, 
we confirm that $\delta\rho(X) \ll \bar\rho$ 
within the mixed regions of Fig.~\ref{FIG:growth}.

Our exponents apply at regime boundaries, where they are continuous. However
our approach does not determine any logarithmic factors.
[Indeed, logarithmic factors are expected at $n=d$, whenever $J_F$ 
or $\tau_B$ dominates, due to the logarithmic divergence of the  
integral for $F$, Eqn.~(\ref{EQN:F}).] At a boundary
between mixed and segregated regimes, any logarithms present determine the
dominant morphology through the self-consistent approach described
above. Without logarithms, the mixed morphology applies
on the boundary, since $W \sim L$ there. The width $W$ characterizes
the mean-distance to annihilation for particles, and thus there is 
a finite density of anti-particles any finite
multiple of $W$ into a domain. When $W  \sim L$, there is a finite
density of anti-particles arbitrarily deep within a domain --- i.e. the
system is mixed. This is the case in the diffusion only 
case at $d=4$, which mixes \cite{Bramson91}.

\subsection{The Reaction Zone}
\label{SEC:W}

For reaction-diffusion systems with a segregated morphology,
much progress has been made on the structure and evolution of the reaction zones
between domains \cite{Lee94,Galfi88,Cornell93,Bennaim92,Leyvraz91}.  
With long-range interactions, we develop a similar approach
that balances the dominant flux {\em into} the reaction zone 
with the dominant annihilation mechanism {\em within} the reaction zone.
This balance determines the reaction zone width, $W(t)$, and the 
interparticle spacing within the reaction zone, $l_{AB}(t)$. To simplify our 
discussion, we take the density of both species to be uniform 
throughout the reaction zone \cite{nonuniform}.  

When a particle enters the reaction zone, it travels 
a typical distance $W$ before it annihilates --- the ``annihilation
mean-free path''.  For a segregated system, 
$W$ must characterize the width of the reaction zone. 
A reaction-zone that grows less rapidly than $W$ allows almost all particles
to pass through unannihilated. 
On the other hand, a reaction-zone that grows more rapidly than $W$ would 
provide infinite number of annihilation mean free-paths as 
$t \rightarrow \infty$, and would 
not have a mix of particles and anti-particles at the far edge of the 
reaction-zone.  Since slower or faster growth is not self-consistent, 
$W(t)$ characterizes the typical reaction zone width.

In the annihilation time $\tau$, a particle diffuses 
a typical distance $W_D \sim (D \tau)^{1/2}$, 
or it ballistically moves $W_B \sim F \tau$ under an applied field $F$,
where $F$ is given by Table~\ref{TAB:F}.
In any case, a particle must move at least the interparticle spacing 
to annihilate, $W_F \sim l_{AB}$.  The largest of these widths
describes how far a particle travels before annihilation, and so 
characterizes the width of the reaction-zone.

The reaction lifetime, $\tau$, the particle spacing in the reaction-zone, 
$l_{AB}$, and the zone width, $W$, are determined self-consistently.  
Given the dominant flux $J$,
we equate the overall flux density into reaction zones $\dot{N}_{flux}
\sim J L^{d-1}/L^d$ with the rate of annihilation within the reaction
zones $W L^{d-1}/[L^d l_{AB}^d \tau(l_{AB})]$. We choose 
the largest $W$ for the given $\tau$,
and the fastest $\tau$ for the given $l_{AB}$. These
dominate the width and annihilation, respectively.  This provides a
self-consistent solution for $\tau$, $l_{AB}$, and $W$.  

For diffusion-only systems in $d \leq 2$, 
our argument is essentially that of Leyvraz and Redner
\cite{Leyvraz91}. We recover their results of $W \sim l_{AB} \sim
t^{3/8}$, and $t^{1/3}$ in $d=1$ and $2$, respectively.  For $d>2$,
however, we have $W \sim \sqrt{\tau_D} \sim l_{AB}^{d/2}$ 
since random-walks are no longer space-filling, 
and we obtain $l_{AB} \sim t^{5/18}$ and $W \sim t^{5/12}$
in $d=3$, and $l_{AB} \sim l_{AA} \sim 1/4$ and $W \sim L \sim t^{1/2}$
in $d=4$.  Our results coincide with the RG scaling results 
of Lee and Cardy \cite{Lee94}.  We also recover 
their scaling relations with respect to currents, whenever
$\tau_D$ dominates annihilation.
In particular, with $\partial_t \bar\rho \sim J/L \sim W/( L l_{AB}^d \tau_D)$,
$\tau_D$ from Eqn.~(\ref{EQN:tauD}), and $W_D \sim \sqrt{\tau_D}$, we have
$W_D \sim J^{-1/3}$ and $l_{AB} \sim J^{-2/(3d)}$ for $d>2$ and 
$W_D \sim l_{AB} \sim J^{-1/(d+1)}$ for $d<2$.   These relations hold
independently of the process that dominates the currents. 

Even within the mixed morphology, it is interesting to consider the
distance $W$ a particle travels in a lifetime $\tau$.  
This distance $W(t)$ is the same scale as the 
remaining charge fluctuations, $X(t)$, as 
discussed before in the previous section.  This is reasonable, since
charge fluctuations at scales much larger than $W$ cannot be flattened out
by charge motion.  We summarize our results in Tab.~\ref{TAB:tau}.

\subsection{Discussion}
\label{SEC:discussion}

In all cases the density decays more slowly as
$n$ increases and the potential becomes sharper and hence shorter ranged.  
If noise is present, then diffusive processes eventually dominate
as $n$ increases. As the spatial dimension $d$ is
increased for any $n$, then a mixed morphology is eventually reached. 
For generalized Coulombic systems, with $n=d-1$, the
density decays as $\bar\rho \sim 1/t$ in all combinations of
high-temperature or uncorrelated initial conditions and force and/or diffusive
processes.  They are always of mixed morphology.

With equilibrated initial conditions, diffusive processes
dominate independently of $n$ when they are present. This is reasonable, 
since the high-temperature initial conditions 
balances the interactions with temperature. Temperature then becomes
more relevant as it is reduced during the quench. 

For systems without diffusive processes, the growth regimes and processes
for uncorrelated and high-temperature initial conditions,
in Fig.~\ref{FIG:growth} a) and d) respectively, correspond.  The various growth
exponents differ only due to the different initial charge fluctuations,
characterized by $\mu$.  On the other hand, with both 
diffusive and long-range processes the regimes
do not correspond, between Fig.~\ref{FIG:growth} b) and e),
because the competition between force and diffusive
processes depends on the charge fluctuations through $\mu$. 

Exponents are continuous at regime boundaries for $l_{AA}$, $l_{AB}$, $L$, 
$\tau$, and $W$.  For segregated systems 
$L \gtrsim W \gtrsim l_{AB} \gtrsim l_{AA}$, where the last
inequality holds since reactions should decrease the particle density.  
We also check that $\tau \lesssim t$ in all cases. 
In $d=1$, we check $W \sim l_{AB}$, 
as expected since the reaction zone is precisely one AB pair. 
At the border between mixed and segregated morphologies,  
the reaction zone is maximal, i.e. $W \sim L$ and $l_{AB} \sim l_{AA}$. 

The particle density at the edge of a domain, 
at $r \sim W$ in Eqn.~(\ref{EQN:domain}), scales the same as
the particle density within the reaction zone, $l_{AB}^{-d}$. i.e. 
that $l_{AB}^{-d} \sim l_{AA}^{-d} (W/L)^{\alpha}$.  Interesting special cases
occur where $J_F$ dominates the flux but $W_D$ dominates the reaction
zone width (regions $[ii^4]$ and $[ii^5]$).
In these special cases, the singular domain profile, with $\alpha <1$, 
leads to a diverging diffusion flux as the domain edge is approached, which must
eventually dominate at some distance $L_X$ from the domain edge. 
In particular, $L_X$ is determined by $\rho(L_X)/L_X \approx J_F$.
Using the domain profile, Eqn.~(\ref{EQN:domain}), 
we  have $L^{-\mu-\alpha} L_X^{\alpha-1} \sim J_F$. 
If $L_X \ll W$, then the cross-over is pre-empted by the reaction zone and
we do not expect to observe it \cite{boundarylayer}. 
However, precisely when $J_F$ and $W_D$ dominate we find that $W_D \ll L_X
\ll L$, so that a linear transition regime is expected for $W_D \ll r \ll L_X$.
The intermediate linear regime, when it occurs, 
does not change the density evolution or reaction-zone structure, 
since $J_F$ still characterizes the flux.

In reaction-diffusion systems, $d_c$ is the critical dimension above which a
coarse-grained reaction rate $\lambda \rho_A \rho_B$ applies, and $d_{u}$ 
can be defined as the dimension above which we can neglect 
inhomogeneities.  Below $d_c$ the reaction term is
not given by $\rho^2$ in a coarse-grained description, and local density
fluctuations must be taken into account \cite{Lee94,Howard95}.  
Above $d_{u}$ we can ignore spatial gradients 
and have $\partial_t \bar\rho \sim - \bar\rho^2$ 
and hence $\bar\rho \sim 1/t$ (see \cite{Ovchinnikov78}).  
Both of these effects stem from the diffusive annihilation mechanism.
The local annihilation rate $\rho/\tau_D$ is proportional to $\rho^2$ 
only for $d > 2$ where $\tau_D \sim \rho^{-1}$. 
This sets $d_c=2$.  The mixed state
is found for $d \geq 4$, so that $d_{u}=4$. 

When only long-range interactions are included, these definitions are not
as useful since neither $\tau_F$ nor $\tau_B$ are in general proportional
to $\rho^{-1}$.  As an example, when only force-driven evolution is included,
as in Fig.~\ref{FIG:growth} a) and d), the density in the {\em mixed} 
state has a $d$-dependent exponent, see $[iii]$ and $[vii]$.
Even the long-wavelength charge fluctuations remain relevant to
drive the dominant ballistic annihilation. 
We could describe this as $d_c = d_{u} = \infty$.

When diffusive and long-range processes are included, however, we see from 
Tab.~\ref{TAB:tau} that 
the diffusive process $\tau_D \sim l_{AB}^d$ dominates above $d_c=2$,
so that the critical dimension is unchanged from the diffusion-only
case. In mixed morphologies diffusive annihilation always dominates,
so that $d_{u}$ equals the dimension where the mixed morphology starts. 
For uncorrelated initial conditions 
we find $d_{u} = n+1$ for $n<3$, while $d_u=4$ above. 
For high-temperature initial conditions $d_{u} = (n+3)/2$ for $n<5$, while
$d_u=4$ above. Since diffusive processes dominate all 
regimes in Fig.~\ref{FIG:growth} e), it is only through the suppression
of initial charge fluctuations that the long-range interactions modify 
$d_u$.

We have assumed that the domains of the segregated
morphology are characterized only by a size $L$ and characteristic
particle separation $l_{AA}$, and that the reaction zones are
similarly characterized by a width $W$ and a particle separation
$l_{AB}$. This leads
to the exponents summarized in the tables and figures. 
It remains possible that other lengths enter into the asymptotic
evolution, for instance through fluctuations in the shape of 
domain boundaries (see, e.g., \cite{Howard95}).  
Our results are most robust for $n<d$, where only the domain
scale $L$ enters in the calculation of the 
forces and currents. Because of that,
we obtain the same result for the evolving density in a segregated morphology 
after a coarse graining to $O(L)$, and any new but shorter lengths 
would not change this result. It remains possible, however, that additional
lengths change the structure of the reaction zone, and hence change
the boundary between segregated and mixed morphologies. Similarly, higher
point correlation effects may lead to more intermediate morphologies
within what we label the mixed regime.
For $n>d$ additional lengths would affect the evolution of the
density, since short scales enter into the 
calculation of the characteristic forces and currents.

Spatial fluctuations of coarse-grained quantities have also been neglected in
our treatment.  We {\em assume} that 
fluctuation effects are not strong enough to change our results for exponents.
Another way to view fluctuations is in terms of the distribution of 
various quantities. We effectively assume that distributions have negligible
tails.

We have also not included the motion of
domain boundaries, either from local heterogeneities in domain density, 
or from differing mobilities of the particle species.  
Domain motion should not affect the scaling exponents {\em if}
the boundaries move slowly enough for the domain profile $\alpha$ to
maintain itself. A simple check is reassuring. The flux required to 
move domain interfaces, $J_{motion} \sim \bar\rho \dot{L} \sim \bar\rho L/t$, 
has the same scaling behavior as the characteristic flux, 
$J \sim L \partial_t{\bar\rho}$. Hence 
domain wall motion should introduce no new scales.

\section{Previous Long-Range Work} 
\label{SEC:prev}

Some of the earliest work on charged $A+B \rightarrow \empty$ systems 
with long-range interactions was by Toyoki \cite{Toyoki88}. He treated
uncorrelated initial conditions and force-only evolution --- 
corresponding to Fig.~\ref{FIG:growth} a).  He presented the mean field
analysis of the mixed morphology, our region $[iii]$.  By considering the 
mean-square force on a particle, he recovered the system-size dependence
for $n<d/2$.  He also initiated numerical studies of $d=2$ systems.
However he used a short-scale cutoff 
proportional to the average particle separation $l_{AA}$
(see Sec.~\ref{SEC:cutoff}), which hampers interpretation of his results. 
Jang {\em et al.} \cite{Jang95} numerically studied
$d=2$ systems with $n=1$ and  obtained 
$\bar\rho \sim t^{-0.55}$ using a large noise amplitude, and $\bar\rho \sim
t^{-0.90}$ using a large force amplitude. This is consistent with our
results of $\bar\rho \sim t^{-1/2}$ and $\bar\rho \sim t^{-1}$ for the pure
noise or pure force systems, respectively.

Ispolatov and Krapivsky \cite{Ispolatov96} focused on 
force-only evolution in $d=1$, with a segregated morphology.  They 
obtained results consistent with ours for $n<d$, where their
assumption of a constant domain profile, with $\alpha=0$, is valid (see 
Table~\ref{TAB:F} and Sec.~\ref{SEC:numerical}). 

Ginzburg, Radzihovsky, and Clark \cite{Ginzburg97} treat 
coarse-grained hydrodynamic systems with uncorrelated
initial conditions and with both diffusion and long-range forces,
for $n \geq d-1$.  Our particle-based treatment can be seen as
complementary to their work. We obtain the same results for density
evolution, but we also obtain details about the reaction zone --- including
$W$, $l_{AB}$, and $\tau$. We also identify the mixed
or segregated morphology of the system.
Furthermore, the system-size dependence in the force for $n<d/2$ 
\cite{Toyoki88,Ispolatov96} is missed in their treatment.  
It would be interesting to extend their approach to include the
particle nature of the charges, possibly with new 
phenomenological annihilation terms. 
Burlatsky {\em et al.} \cite{Burlatsky96} present a simple scaling
model that obtains the same results as the self-consistent
approach of Ginzburg {\em et al.} \cite{Ginzburg97}.

\section{Electronic Systems}
\label{SEC:electrical}

Asymptotic non-geminate pair-recombination in clean
crystalline semiconductor systems 
should be described by our approach with $n=2$ and
equilibrated high-temperature initial conditions.  
In $d=3$, we find $\bar\rho \sim t^{-1}$ with a mixed morphology. This result 
applies for every combination of uncorrelated or high-temperature initial
conditions and force and/or diffusion. This is because {\em all} 
annihilation mechanisms have the same scaling, $\tau \sim l_{AB}^3$. 

In contrast, for $d<3$ the annihilation mechanisms differ and the
morphologies are all {\em segregated}. The structure of the 
evolving charge-fluctuations is thus much richer.
Specifically, in $d=1$, and $2$ we expect regime $[viii']$ to apply 
asymptotically with $\bar\rho \sim t^{-1/4}$ and $t^{-3/4}$ respectively. 
The decay rate is dramatically slowed due to the segregated morphology.

Clean $d=2$ systems exist in quantum hall effect (QHE) devices.  
QHE devices with skyrmion charge excitations (see, e.g., \cite{Barrett95})
may be particularly good systems to study these effects, due to
their low mobilities and slow dynamics combined with sensitive 
time-resolved probes of their particle density. 
We will develop this in more detail in a separate
publication \cite{Rutenberg98}, 
paying particular attention to the scaling of the amplitudes 
and the resulting pre-asymptotic cross-overs in the evolution.

\section{Particle-Like Topological Defects}
\label{SEC:topo}

Point-like topological defects, e.g. hedgehogs in $d=3$ or vortices in $d=2$, 
are found in liquid crystals and in vector $O(N)$ systems in $d=N$
dimensions. Considered pairwise, these defects have power-law interactions.
Indeed, early theoretical work \cite{Toyoki88} on these systems 
was based on the interactions of point-like topological defects.   
However the evolution of systems with point-like topological defects is
different than the dynamics of Sec.~\ref{SEC:longrange} (see however
\cite{oned}). The order parameter field that supports topological defects  
provides a scale-dependent mobility to particle motion. 
Additionally, the mobility depends on the local environment, i.e. the
interaction between defects is not a two-point particle-particle 
interaction. We may, however, take that as a first approximation. 
For $d=2$ XY or nematic systems, the mobility scales logarithmically
with the particle velocity, $\eta(l) \sim 1/\ln(dl/dt)$ \cite{Rutenberg95}. 
For $d=3$ $O(3)$ or nematic systems, the mobility of hedgehogs scales 
as the inverse separation, $\eta(l) \sim 1/l$ \cite{Pargellis91}. 
In general, for $O(N)$ models in $N \geq 3$ dimensions, $\eta(l) \sim l^{2-N}$
\cite{Rutenberg95}.  High-temperature equilibrated initial conditions are 
appropriate for quenches in physical systems. 
For point defects in $d \geq 3$, the effective
interaction is linear \cite{Toyoki88} ($n=0$), which has not been treated
in this paper.  However, $d=2$ systems are within our purview.

For the $d=2$ XY model, interactions between defects are logarithmic
($n=1$). Ignoring the logarithmic mobility, and including equilibrated
initial conditions \cite{2dxy}, we have a mixed morphology with 
all lengths  scaling the same, $l_{AA} \sim l_{AB} \sim L \sim W \sim t^{1/2}$.
This is found whether or not diffusive processes are included, and indeed
$\tau_D \sim \tau_B \sim \tau_F \sim l_{AB}^2$ and $J_D \sim J_F \sim
L^{-3}$ for this system. The single length scale and the similarities
between no noise quenches and quenches with diffusion 
matches the phase ordering results of the 
$d=2$ XY model \cite{Rutenberg95,Yurke93}. 

Two phase-ordering systems that do exhibit strong scaling violations are
the $d=1$ XY model \cite{Rutenberg95ii}
and the $d=2$ $O(3)$ model \cite{Rutenberg97}. Both of these systems 
support non-singular topological textures that have particle-like aspects.
However topological textures have an
intrinsic length-scale that evolves in time. Significant extensions to
our approach would be necessary for these systems. 
We may also consider interacting topological defects in the patterned 
structures of driven (see, e.g., \cite{Cross95}) systems or of systems
with competing interactions (see, e.g., \cite{Sagui95}). 
To apply our approach,  the long-wavelength fluctuations ($\mu$), 
the interaction ($n$), the mobility
$\eta(l)$, and the nature of any noise-driven transport must be identified. 

\section{L\'{e}vy Super-diffusion}
\label{SEC:levy}

Systems with long-range L\'{e}vy super-diffusion \cite{Shlesinger93} have been
used to model stirred reaction-diffusion systems. 
While there are no long-range interactions {\em per se} in these systems, 
super-diffusion enhances the reaction rate in a
manner qualitatively similar to long-range interactions.  Indeed, our
methods can be applied to this case and agree with the 
results of Zumofen {\em et al.} \cite{Zumofen96}
for the late-time evolution from uncorrelated initial conditions.
We also obtain additional information about the reaction zone structure 
and interface profiles for segregated morphologies. 
For simplicity, we only consider uncorrelated initial conditions, with 
$\mu =d/2$. 

In the discrete formulation of L\'{e}vy flight, every particle hops
a random distance $r$ along a random lattice direction with 
probability distribution $P(r) \propto r^{-1-\gamma}$ --- 
annihilating any anti-particles it encounters along the way. 
We impose $1< \gamma <2$ so that the hop distribution has a finite
first moment and is normalizable. 
The equivalent continuum dynamics is 
$\partial_t{\rho}_k = - D |k|^\gamma \rho_k$, where $\gamma>1$ is required.

In time $t$, taken to be large, a particle randomly L\'{e}vy 
walks a distance $R \sim t^{1/\gamma}$. If $\gamma>d$, the volume bounding
this walk $R^d \ll t$ and the walk is recurrent, or space-filling. 
Conversely, if $\gamma<d$ the walk is sparse, and the particle explores 
a volume proportional to $t$ using its finite capture radius $r_c$. 
Paralleling Eqn.~(\ref{EQN:tauD}), the 
annihilation time in a well-mixed region of the system scales as
\begin{equation}
\label{EQN:tauL}
	\tau_L \sim  \left\{
                \begin{array}{c}
  		l_{AB}^{\displaystyle d}           \ \ \ \ \ \    \gamma < d, \\
       		 l_{AB}^{\displaystyle \gamma}         \ \ \ \ \ \  	\gamma > d, 
                \end{array} \right. 
\end{equation}
where the typical particle-antiparticle separation in the region is
$l_{AB}$. From the L\'{e}vy flight, 
the net distance that the particles move before 
annihilating is $W \sim \tau_L^{1/\gamma}$. For space-filling walks
($\gamma>d$) $W\sim l_{AB}$, while for sparse walks ($\gamma<d$) 
$W \sim l_{AB}^{d/\gamma} \gg l_{AB}$. 

For a segregated morphology, there is a typical flux-density 
leaving domains. 
The flux-density is most easily obtained in the discrete formulation, where it
is simply the number of particles that pass through a site in each
time-step. From the L\'{e}vy flight distribution, a fraction
$p \sim 1/x^\gamma$ of particles contributes to the current from a given
distance $x$ away.  Only particles along lattice
directions contribute in a single time-step and, since $\gamma >1$, 
the flux is dominated by particles nearby compared
to the domain scale.
Hence, our current is $J(r) \sim \int_0^r dy 
[\rho(r+y)- \rho(r-y)]/y^\gamma$ a distance $r$ from a domain edge,
where $W \ll r \ll L$. 
Using the domain profile from Eqn.~(\ref{EQN:domain}),
and requiring that $J(r)$ approaches a constant near
the domain edge we obtain a non-linear 
profile exponent, $\alpha=\gamma-1$, and
$J_L \sim \bar\rho L^{1-\gamma}$ for the particle current. 

For segregated systems, we equate the flux out of domains 
$\partial_t{\bar\rho} \sim J L^{d-1}/L^d \sim l_{AA}^{-(d+2\gamma)}$ 
to the time derivative of the charge density $l_{AA}^{-d}/t$ to obtain
$l_{AA} \sim t^{1/(2\gamma)}$ and $\bar\rho \sim t^{-d/(2\gamma)}$. The
domain size $L \sim t^{1/\gamma}$.  We can also obtain the rate of particle
annihilation from the reaction rate within the reaction zones,
$\partial_t{\bar\rho} \sim W L^{d-1}/[L^d l_{AB}^d \tau_L]$, and
compare to obtain
\begin{equation}
\label{EQN:labL}
	l_{AB} \sim  \left\{
                \begin{array}{c}
        		 t^{(d+2\gamma-2)/[2 d(2 \gamma-1)]}\ \ \ \ \ \ \ \gamma <d,\\
        		 t^{(d+2\gamma-2)/[2 \gamma(\gamma+d-1)]}  
												 \ \ \ \ \ 	\gamma >d .
                \end{array} \right.
\end{equation}
Since particles are randomly L\'{e}vy-walking before annihilation, the 
reaction zone width is $W \sim l_{AB}$ for $d<\gamma$, and 
$W \sim l_{AB}^{d/\gamma}$ for $d>\gamma$. 

The system has a segregated morphology,
with $W \lesssim L \sim l_{AA}^2$, for $\gamma>d/2$. For $\gamma<d/2$ the 
system is mixed, and the time-scale is set by the annihilation rate
$\tau_L \sim t$, so that $l_{AA} \sim t^{1/d}$ and $\bar\rho \sim 1/t$.  
At $\gamma=d/2$, the marginal case between segregated and mixed morphologies,
we have $W \sim L$ and $l_{AB} \sim l_{AA}$.
Our results agree with those of Zumofen {\em et al.} \cite{Zumofen96}.
We also obtain $\alpha$, $l_{AB}$, and $W$ characterizing the domain
profile and reaction zone in segregated systems.  

It is interesting that while our results are qualitatively similar to 
those obtained with long-range {\em interactions}, there is no effective 
interaction $n$ that recovers the growth-exponents for a given L\'{e}vy exponent
$\gamma$.  This is in contrast to the nonequilibrium steady-state
properties of the kinetic Ising model, 
where L\'{e}vy flights generate effective long-range interactions
\cite{Bergersen91}.

We may be able to apply these results to the motion of an single charged 
particle in a quenched random potential.  This motion 
is sub-diffusive, with $\gamma>2$ and $R \sim t^{1/\gamma}$, for potentials
with sufficiently long-range correlations \cite{Bouchaud90}.  
One loop RG calculations by Deem and Park \cite{Deem98} 
for potentials caused by quenched Coulombic ions in $d=2$ ($n=1$),
matched to the homogeneous evolution without any long-range interactions,
indicate that sub-diffusive behavior alone may be sufficient to 
describe the evolution of the quenched system.  If so, then 
our L\'{e}vy-flight results may be directly applied with the 
appropriate $\gamma$.  However significant questions remain,
such as the appropriate microscopic annihilation mechanisms for oppositely 
charged particles in a quenched random potential.

\section{Long-Range Cutoffs}
\label{SEC:cutoff}

It is interesting to explore quenched long-range systems numerically,
but the computational burden can be large. Long-range cutoffs $L_{cut}(t)$,
usually on the order of the interparticle spacing, 
accelerate a simulation, but often at the cost of changing the physics. 

Long-range cutoffs contribute in two places. The first
is through the force integral driving particles fluxes, 
$F(r)$ in Eqn.~(\ref{EQN:F}).
Any $L_{cut} \ll L(t)$ replaces the upper-cutoff of the integral
and change $F(r)$ for the cases where the integral is not dominated
by short scales, i.e. for $n<d+1$.   If $F(r)$ is changed, the
domain profile exponents $\alpha$ are also affected. 
The second place cutoffs enter 
is through the annihilation time within the reaction zone, 
Eqn.~(\ref{EQN:tauF}).  If the cutoff is smaller than the reaction-zone
spacing, $L_{cut} \ll l_{AB}(t)$, then the annihilation
dynamics of $\tau_F$ is changed.  Ballistic annihilation
times $\tau_B$ is qualitatively changed whenever $L_{cut} \ll L(t)$. 
Amplitudes can also be affected by cutoffs, even when growth exponents are not. 

As an illustration, we consider limiting interactions to  
nearest-neighbors.  Consider
uncorrelated initial conditions with no diffusion, Fig.~\ref{FIG:growth} a). 
For nearest-neighbor interactions, we 
use $L_{cut} \sim l_{AA}(r) \sim \rho(r)^{-1/d}$, the local 
particle spacing. Putting $L_{cut}$ instead of $L$ in
Eqn.~(\ref{EQN:F}) leads to the same results as $n>d+1$ in Table~\ref{TAB:F}
for {\em all} $n$.  In particular, the reaction-zone profile is
characterized by $\alpha = d/(n+d-1)$. 
The nearest-neighbor cutoff is of order $l_{AB}$  within 
the reaction zone so that local force-driven  
annihilation dynamics, with $\tau_F$, 
is qualitatively unchanged.  The system size cannot enter the force
integral, so there is no restriction on the interaction to $n>d/2$. 
For a segregated morphology regime $[i]$  applies, while in the mixed
regime $[iii]$ does,
though with $W \sim l_{AB}$ and annihilation through $\tau_F$.
Comparing the particle annihilation rate $\partial_t{\bar\rho} \sim J/L$ to 
the maximum rate supported by reaction zones
$\partial_t{\bar\rho}_{max} \sim \bar\rho/\tau_F(l_{AA})\sim 1/l_{AA}^{n+d+1}$,
we find that segregation occurs for all $d<3$. In summary,
nearest neighbor interactions only leave the evolution qualitatively unchanged
for $n>d+1$. A numerical test of the effects of nearest-neighbor interactions 
in $d=1$ is made in the next section.

\section{Numerical Investigation}
\label{SEC:numerical}

We present some numerical results on the non-trivial domain profile 
exponent $\alpha$ in uncorrelated one-dimensional systems without diffusion. 
We also consider the effect of a cutoff $L_{cut}$ limiting interactions to 
nearest-neighbors. More results will be presented in future work
\cite{Rutenberg98}.  

In $d=1$, we expect $\alpha= (n-1)/2$ for $1<n<2$ (regime $[ii^0]$) 
and $\alpha=1/n$ for $n>2$ (regime $[i]$). 
We studied systems with $8000$
particles with $n=3/2$ [$375$ samples] and $n=2$ [$300$ samples].  
We expect $\alpha = 1/4$ and $1/2$ respectively. 
We avoided finite-size effects
by comparisons with size $4000$ systems. 
We also considered systems with only nearest-neighbor interactions, which
are expected to modify the domain profile
for $n<2$, see Sec.~\ref{SEC:cutoff}. In particular, we expect regime $[i]$
to apply for {\em all} $n<2$. For $n=3/2$, we expect $\alpha = 2/3$.  
We studied systems with $128000$ particles with $n=3/2$ [$200$ samples] and 
$n=2$ [$53$ samples], and checked finite-size effects with systems of $32000$ 
particles. We found results consistent with expectations. We show our
results for the domain profiles in Fig.~\ref{FIG:prof}.


\section{Conclusions}
\label{SEC:summary}

\begin{figure}
\begin{center}
\mbox{
\epsfxsize=17.0cm
\epsfbox{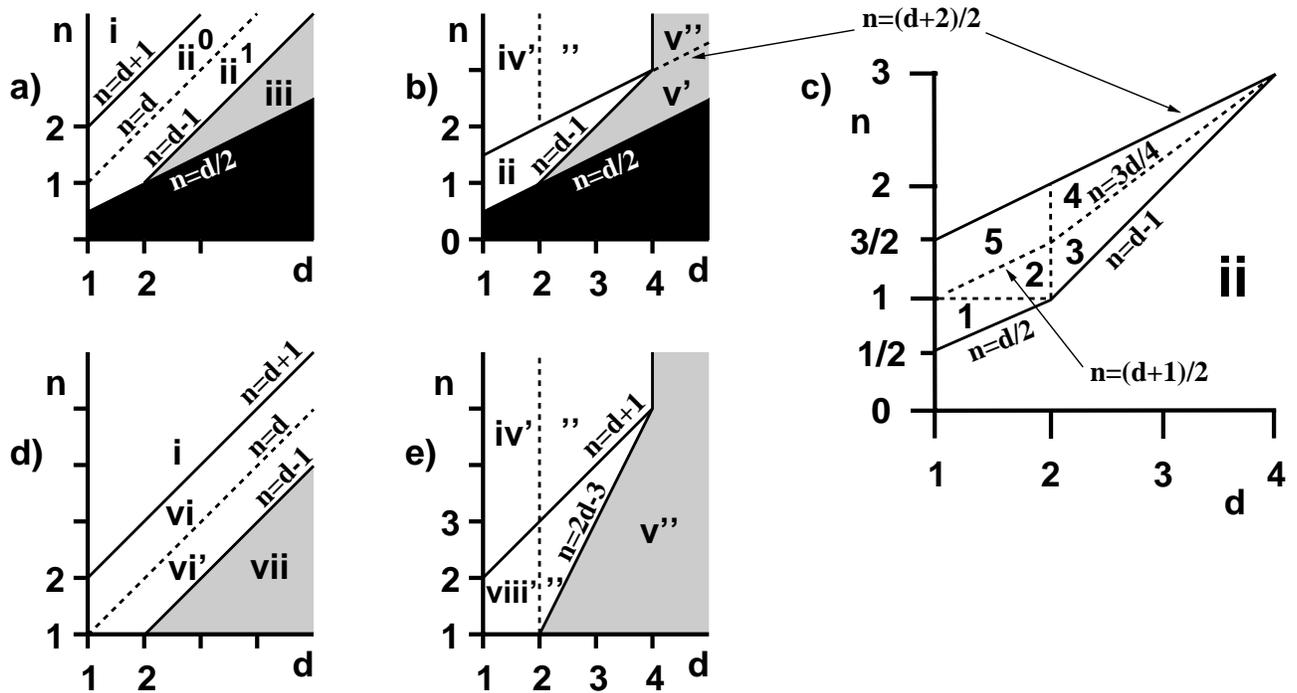}}
\end{center}
\caption{Different regimes of density evolution.  Blank regions
indicate a segregated morphology, shaded a mixed morphology, while black 
indicates excluded regimes that have no thermodynamic limit. 
Dashed lines separate regimes, numbered
with Roman numerals, with different reaction-zone structure (see
Table~\protect\ref{TAB:tau}).  The figures correspond to: 
a) uncorrelated initial conditions, force only, 
b) uncorrelated initial conditions, diffusion and force, 
c) enlargement of b), with the $[ii^1]$ to $[ii^5]$ subregions numbered, 
d) equilibrated high-temperature initial conditions, force only, 
and e) equilibrated initial conditions, noise and force.  
\label{FIG:growth}}
\end{figure}

By considering the annihilation dynamics in well mixed regions of a
charged particle system, and balancing the annihilation against currents driven
by charge inhomogeneities left over from the initial conditions, we
self-consistently determine the morphology and evolution
of quenched charged-particle systems with long-range interactions. 
We also contribute a visceral description of the dynamics.
We characterize the system with the scale of domains $L$ and reaction-zones
$W$, and the particle spacings within domains $l_{AA}$ and reaction-zones
$l_{AB}$.  For mixed systems $L \sim l_{AA} \sim l_{AB}$. 
Our results are summarized in Fig.~\ref{FIG:growth}, and in the three tables.
Our primary assumption is that the lengths we have used are sufficient
to characterize the evolving system. The scaling form for the domain
profile, Eqn.~(\ref{EQN:domain}), follows from this assumption.

The results of this paper will hopefully inspire more formal derivations
and numerical tests, as well as experimental tests in electronic systems. 
Comparisons with existing treatments is encouraging,
particularly agreement with the field-theory approach of Lee and Cardy
\cite{Lee94} for reaction-diffusion systems, with the hydrodynamic treatment of 
Ginzburg, Radzihovsky, and Clark \cite{Ginzburg97} for uncorrelated
initial conditions with diffusion and long-range interactions, and with
the L\'{e}vy super-diffusion results of Zumofen {\em et al.}
\cite{Zumofen96}.  There are also many new results pertaining 
to the reaction-zones, domain profiles, and systems with equilibrated
high-temperature initial conditions. 

Electronic systems ($n=2$) in two-dimensions should provide an experimental
test for our results. Asymptotically, we predict a segregated morphology
with domain size $L \sim t^{1/2}$ and average density 
$\bar{\rho} \sim t^{-3/4}$ for photo-excited 
quantum-well or QHE systems (high-temperature equilibrium initial conditions, 
regime $[viii'/viii'']$).  This will be explored at more length in
a separate publication \cite{Rutenberg98}.  
If initial conditions with uncorrelated
initial conditions can be manufactured, then we expect regime
$[iv'/iv''/ii^4/ii^5]$ to apply: a segregated morphology with $L \sim t^{1/2}$
and $\bar{\rho} \sim t^{-1/2}$.  These contrast dramatically with the
mixed morphology and density decay $\bar{\rho} \sim t^{-1}$ in
three-dimensional electronic systems. 

I thank the EPSRC for support under Grant No. GR/J78044, 
the NSERC, and {\it le Fonds pour la Formation
de Chercheurs et l'Aide \`a la Recherche du Qu\'ebec}.
I also thank John Chalker, Slava Ispolatov, Ben Lee, Klaus Oerding,
Zoltan R\'{a}cz, and Beate Schmittmann for useful discussions.

\newpage
\onecolumn
\pagebreak
\begin{table}
\begin{tabular}{|l|ccc|} 
        	& $\alpha$ & $F(r)$ & $J_F$  \\ \hline
$n<d$       & $0$ & $L^{d-n-\mu}$ & $L^{d-n-2\mu}$ \\
$d<n<d+1$   & $(n-d)/2$ & $r^{(d-n)/2} L^{(d-n-2\mu)/2}$ & $L^{d-n-2\mu}$ \\
$n>d+1$		& $d/(n+d-1)$ & $(L/r)^{d/(n+d-1)} L^{-(n+1)/2}$ 
						  & $L^{-(n+d+1)/2}$ \\
\end{tabular}
\caption{Domain profile exponents, $\alpha$, coarse-grained 
field a distance $r \ll L$ from the domain edge, $F(r)$, and 
force-driven flux, $J_F$, for different interaction exponents $n$. 
When diffusive fluxes dominate $\alpha=1$.}
\label{TAB:F}
\end{table}
\nopagebreak[4]
\begin{table}
\begin{tabular}{|l|ccccc@{\hspace{0.1in}}|} 
        	& $d \ln{l_{AA}}/d \ln{t}$ & $d \ln{L}/d\ln{t}$ & $d\ln{l_{AB}}/d
			\ln{t}$ & $ - d \ln{J}/d\ln{t} $ &J \\ \hline
i       & $1/(n+3)$	
        & $2/(n+3)$
		& $(n+d+1)/[(n+d)(n+3)]$
		& $(n+d+1)/(n+3)$ 
        & F \\
\hline
ii$^0$  & $1/(2n+2-d)$
        & $2/(2n+2-d)$ 
		& $2n/[(n+d)(2n+2-d)]$
		& $2n/(2n+2-d)$ 
        & F \\
ii$^1$  & "
        & "
		& $l_{AA}$
		& " 
        & F \\
ii$^2$  & "		
        & "
		& "
		& " 
        & F \\
ii$^3$  & "		
        & "
		& "
		& " 
        & F \\
ii$^4$  & "		
        & " 
		& ${4n}/{[3d(2n+2-d)]}$
		& "
        & F \\
ii$^5$	& "		
        & "
		& $2n/[(d+1)(2n+2-d)]$
		& " 
        & F \\
\hline
iii     & $(2n+1-d)/[n(2n+2-d)]$ 
        & $l_{AA}$
		& $l_{AA}$
		& " 	
        & F \\
\hline
iv$'$   & $1/4$
        & $1/2$ 
		& ${(d+2)}/{[4(d+1)]}$
		& $(d+2)/4$  
        & D \\
iv$''$  & "
        & "
		& $(d+2)/(6d)$
		& " 
        & D \\
\hline
v$'$    & ${1}/{d}$
        & $l_{AA}$
		& $l_{AA}$
		& $2n/(2n+2-d)$	
        & F \\
v$''$   &  "
        & "
		& "
		& $(\mu+1)/2$	
        & D \\
\hline
vi      & $(2d+1-n)/[d (n+3)]$	
        & $2/(n+3)$
		& $2(d+1)/[(d+n)(n+3)]$
		& $2 (d+1)/(n+3)$ 
        & F \\
vi$'$   &  "
        &  "
		& $l_{AA}$
		& " 
        & F \\
\hline
vii     &  $(n+nd+d-1)/[nd(n+3)]$
        & $l_{AA}$
		& $l_{AA}$
		& " 
        & F \\
\hline
viii$'$ & ${\mu}/(2d)$	
        &  $1/2$
		& $(\mu+1)/[2(d+1)]$
		& $(\mu+1)/2$
        & D \\
viii$''$& "	 	
        & "
		& $(\mu+1)/(3d)$
		& " 
        & D \\
\end{tabular}
\caption{Growth laws of $l_{AA}$, $L$, $l_{AB}$, and $J$ for 
the different labeled regimes in Fig.~\protect\ref{FIG:growth}. Lengths
with the same scaling as $l_{AA}$ are indicated. The
last column indicates the dominant flux mechanism, where
``D'' is diffusive and ``F'' is force driven. }
\label{TAB:growth}
\end{table}
\nopagebreak[4]
\begin{table}
\begin{tabular}{|l|ccccc|} 
        	& $W$ & $d \ln{W}/d \ln{t}$ & $\tau$ & $d \ln{\tau}/d\ln{t}$& 
				$d \ln{l_\ast}/d \ln{t}$  \\ \hline
i      	& F
		& $l_{AB}$
		& F
		& $(n+1)(n+d+1)/[(n+d)(n+3)]$
		&  --- \\
\hline
ii$^0$  & F
		& $l_{AB}$
		& F
		& $2n(n+1)/[(n+d)(2n+2-d)]$
		& --- \\
ii$^1$  & B  
		& ${(d^2+2n-nd-d)}/{[n(2n+2-d)]}$
		& B
		& $1+d(d-n-1)/[n(2n+2-d)]$
		& $(2n-d)/[n(2n+2-d)]$ \\
ii$^2$  & B
		& $(d^2+2n-2nd)/(2n+2-d)$
		& B
		& $(d^2+4n-2nd-d)/(2n+2-d)$ 
		& " \\ 
ii$^3$  & B
		& ${2(d-n)}/{(2n+2-d)}$
		& D 
		& $d/(2n+2-d)$ 
		& ---  \\
ii$^4$  & D
		& ${2n}/{[3(2n+2-d)]}$
		& D 
		& $4n/[3(2n+2-d)]$
		& ---  \\
ii$^5$	& D
		&  $l_{AB}$
		& D
		& $4n/[(d+1)(2n+2-d)]$ 
		& ---  \\
\hline
iii     & B
		& $2/(2n+2-d)$
		& B
		& $1$ 	
		& $(2n-d)/[n(2n+2-d)]$ \\
\hline
iv$'$   & D
		& $l_{AB}$
		& D  
		& $(d+2)/[2(d+1)]$ 
		& --- \\
iv$''$  & D
		& $(d+2)/{12}$
		& D 
		& $(d+2)/6$ 
		& --- \\
\hline
v$'$    & B
		& $2/(2n+2-d)$ 
		& D  
		& $1$ 	
		& --- \\
v$''$   & D
		&  $1/2$
		& D  
		& $1$ 	
		& --- \\
\hline
vi      & F
		& $l_{AB}$
		& F 
		& $2(d+1)(n+1)/[(d+n)(n+3)]$ 
		& --- \\
vi$'$   & B
		& $(2n+nd+1-n^2-d)/[n(n+3)]$
		& B  
		& $(3n+nd+1-d)/[n(n+3)]$
		& $(n+1)/[n(n+3)]$ \\
\hline
vii     & B
		& $2/(n+3)$
		& B  
		& $1$
		& " \\
\hline
viii$'$ & D
		&  $l_{AB}$
		& D  
		& $(\mu+1)/(d+1)$ 
		& --- \\
viii$''$& D
		& $(\mu+1)/{6}$
		& D 
		& $(\mu+1)/3$
		& --- \\
\end{tabular}
\caption{Growth laws of $W$, $\tau$, and $l_\ast$ for 
the different labeled regimes in Fig.~\protect\ref{FIG:growth}. Reaction-zone
widths $W$ with the same scaling as $l_{AB}$ are indicated. The
dominant mechanisms determining $W$ and $\tau$ are also indicated, where
``D'' is diffusive, ``F'' is force driven, and ``B'' is ballistic. Note
that when the morphology is mixed, $\tau \sim t$ and $W$
indicates the scale of remaining charge fluctuations.}
\label{TAB:tau}
\end{table}
\twocolumn


\begin{references}
\bibitem{propagator} No diagrams dress the particle 
propagators of formal approaches when the only interaction is annihilation 
(see \cite{Lee94}). This is not true when long-range interactions are 
included.
\bibitem{immobile} When one species is immobile 
correlations can become important \protect\cite{Leyvraz91}, 
particularly for the domain structure of the 
immobile species.  For purely diffusive systems,
this is a singular limit \protect\cite{Lee94}.
\bibitem{equil} There is no pair-creation, 
$\emptyset \rightarrow A+B$, in our dynamics. 
This corresponds to being at a temperature far below 
the chemical potential of pairs.
\bibitem{initialdensity} The initial 
particle density is tuned by a chemical potential. 
It does not affect the asymptotic exponents of the subsequent evolution.
\bibitem{2dxy} A quenched 2d XY model can be approximated by a system 
of charged-vortices interacting logarithmically, with $n=1$ 
(see Sec.~\ref{SEC:topo}).
Coarse-grained charge fluctuations, as described by Eqn.~(\ref{EQN:rhok}),
lead to the correct mixed morphology with $L \sim l_{AA} \sim t^{1/2}$ 
(up to logarithms, see, e.g., \protect\cite{Rutenberg95,Yurke93}).
In contrast, characterizing the initial fluctuations by 
the excess number of vortices within
a sharply defined box of size $L$, i.e. $\mu=3/2$,
leads to a segregated morphology with
$L \sim t^{1/2}$ but $l_{AA} \sim t^{3/8}$ \cite{Mondello90},
which is not observed \protect\cite{Rutenberg95,Yurke93}.
\bibitem{fractalprofile} We coarse-grain the charge density to a small but
finite fraction of $L$, and we assume a smooth domain structure at that scale. 
Alternatively, we could build in a fractal domain structure 
(see \cite{Leyvraz91}), but we cannot a priori determine if a fractal 
structure evolves from the dynamics.
\bibitem{selfconsist} We self-consistently determine the shortest applicable 
annihilation time $\tau(l_{AB})$, the largest relevant reaction zone width 
$W(\tau)$, and the corresponding particle spacing in the reaction-zone 
$l_{AB}$, in terms of their time-exponents. The results are given 
in Tables~\protect\ref{TAB:growth} and \protect\ref{TAB:tau}.
\bibitem{othermorphologies} Other morphologies are possible in principle,
though they have not been observed in reaction-diffusion systems. 
They would introduce new length-scales.
\bibitem{nonuniform} In general, reaction zones are inhomogeneous,
with profiles scaled by the reaction zone width $W$ and particle-spacing
$l_{AB}$ \protect\cite{Galfi88,Bennaim92}, and with dilute multi-scaling
tails \protect\cite{Howard95}.  Our analysis is unchanged by these refinements.
\bibitem{boundarylayer} This linear transition zone is precisely the 
``boundary layer'' in the mean-field treatment of Ben-Naim and Redner
\protect\cite{Bennaim92}. Note that force-driven and ballistic annihilation 
mechanisms, not contained in their approach, 
stabilize the reaction zone when $L_X \ll W$.
\end{references}
\end{document}